\documentclass[aps,twocolumn,showpacs,superscriptaddress,amsmath]{revtex4}
\usepackage{amsmath,amsfonts}
\usepackage{epsfig}
\usepackage{graphicx}

\newcommand{\bea}{\begin{eqnarray}}
\newcommand{\beq}{\begin{equation}}
\newcommand{\eea}{\end{eqnarray}}
\newcommand{\eeq}{\end{equation}}
\begin{document}
\title
{Entanglement control in coupled two-mode boson systems}
\author{A.\ V.\ Chizhov}
\affiliation{Bogoliubov Laboratory of Theoretical Physics,
Joint Institute for Nuclear Research, 141980 Dubna, Russia}
\author{R.\ G.\ Nazmitdinov}
\affiliation{Bogoliubov Laboratory of Theoretical Physics,
Joint Institute for Nuclear Research, 141980 Dubna, Russia}
\affiliation{Departament de F{\'\i}sica,
Universitat de les Illes Balears, E-07122 Palma de Mallorca, Spain}


\begin{abstract}
A two-mode boson  model, widely used for the physics of
fast rotating nuclei and Bose-Einstein condensates, is
studied in the context of entanglement control.
We derive an analytical expression for the entanglement between the fields in
this model as a function of time.
We found  that depending on the
interaction strengths between boson modes and
the nature of the initial boson states
the dynamical evolution of the entanglement and
the squeezing can occur independently.
\end{abstract}

\maketitle
There is currently an enormous effort underway to understand the rich
dynamics of interacting Bose systems. Among various problems are the properties
of Bose-Einstein condensate (BEC) in external fields \cite{PS},
the Bose coherent effects of excitons and polaritons in semiconductor
microcavities \cite{kav}, and the production of scattered radiation due to
interaction of the incident laser beam
with vibrational modes of a medium (the Raman effect).
The generic feature of interacting Bose systems is the formation of Bose-field
collective states with nonclassical (squeezed, sub-Poissonian, etc) statistical
and fluctuation properties.
Another prominent feature is an entanglement produced by the
interaction between different system constituents,
which is one of the most subtle and intriguing phenomena
in nature (see for a recent review \cite{am,hor}).
Nowadays, there is  explosive activity in the study of the entanglement
due to its potential usefulness
in quantum teleportation, quantum cryptography, and, in general,
in quantum information theory.

The dynamical interplay between quantum entanglement and nonclassical
properties of various Bose systems can be traced within
a model of two-coupled harmonic oscillators.
A particular example of interest is the Hamiltonian
\bea
\label{eq1}
\hat{H}&=&\hbar\omega_1\hat{c}_1^\dag\hat{c}_1
+\hbar\omega_2\hat{c}_2^\dag\hat{c}_2\nonumber\\
&&+i\hbar g_1\left(\hat{c}_1^\dag\hat{c}_2-\hat{c}_2^\dag\hat{c}_1\right)
-i\hbar g_2\left(\hat{c}_1^\dag\hat{c}_2^\dag-\hat{c}_2\hat{c}_1\right)\,.
\eea
Here $\hat{c}_{1,2}$ ($\hat{c}_{1,2}^\dag$) are the annihilation (creation)
boson operators of the fields with the energies
$\hbar\omega_{1,2}$. The coupling between these fields is governed
by dint of the coupling constants $g_{1,2}$.
Note that the bilinear form of the Hamiltonian (\ref{eq1}) corresponds to a
linearized version of some more general interactions.

This model has been applied in nuclear physics \cite{BM75,BR} and
for a rotating BEC (cf. Ref.~\onlinecite{fet}).
In condensed matter physics the Hamiltonian (\ref{eq1}) is used:
i) to study the interaction between an atom
and a radiative field \cite{sh}; ii) as
a starting point for analysis of
electronic properties of two-dimensional quantum dots in
a perpendicular magnetic field \cite{qd1,qd2}.
The dynamics of two-component Bose condensate trapped in
a double-well potential can be also mapped on the time-evolution
of two coupled-harmonic oscillators in the low excitation regime \cite{ng}.
In the simplest case, the model describes two levels of the condensed
atoms which are coupled owing to the classical field of radiation \cite{las}.
In this case the
interaction constants ($g_1$ and $g_2$) in the Hamiltonian (\ref{eq1})
simulate the coupling between the electromagnetic field and the two-level
(two-mode) boson system. Varying the interaction (the coupling) and the choice
of  initial states one can analyze various properties of this system.

The bilinear form of boson interaction enables one
to investigate the system dynamics analytically.
In particular, the exact solutions as well as the
generation of squeezed states
in this model have been analyzed in \cite{CN90}.
The main aim of the present paper is to gain a better insight
into the dynamical
interplay between the strength of the interaction, entanglement
and squeezing dynamics of the system. Here, we focus upon the dynamics
of Gaussian states, which are of great practical importance.

By means of the Bogoliubov transformation
\begin{equation}
\label{eq2}
\hat{a}_k=\sum_{m=1}^2\left( A_m^k\hat{c}_m + B_m^k\hat{c}_m^\dag
\right)\,,
\end{equation}
the Hamiltonian~(\ref{eq1}) is reduced
to the diagonal form
$\hat{H}=\sum_{k=1}^2\hbar\Omega_k\left(\hat{a}_k^\dag\hat{a}_k +
1/2\right)$
with the two eigenmodes
\begin{eqnarray}
\label{eq3}
\Omega_k^2 &=& [\eta_{+}+2(g_1^2-g_2^2)+
(-1)^{k+1}\Delta]/2\,,\\
\Delta  &=&  [\eta_{-}^2+4\eta_{+}(g_1^2-g_2^2)
+8\omega_1\omega_2(g_1^2+g_2^2)]^{1/2}\,,\nonumber
\end{eqnarray}
where $\eta_{\pm}=\omega_1^2\pm\omega_2^2$.
The coefficients $A_m^k$ and $B_m^k$ of the transformation matrices in
Eq.~(\ref{eq2}) are defined in Ref.~\onlinecite{CN90}.

By virtue of the inverse transformation and
time evolution of the collective
operators $\hat{a}_k(t)=\hat{a}_k(0){\rm e}^{-i\Omega_k t}$,
one can determine the time dependence of the
initial operators $\hat{c}_m$
\begin{equation}
\label{eq6}
\hat{c}_m(t)=\sum_{n=1}^2\left[ \alpha_{mn}(t)\hat{c}_n(0)
+ \beta_{mn}(t)\hat{c}_n^\dag(0) \right] \,,
\end{equation}
where
\begin{eqnarray}
\label{eq7}
\alpha_{mn}(t) &=& \sum_{k=1}^2\left[ A_m^{*k}A_n^{k}{\rm e}^{-i\Omega_k t}
-B_m^{k}B_n^{*k}{\rm e}^{i\Omega_k t} \right]\,,\nonumber\\[1mm]
\beta_{mn}(t) &=& \sum_{k=1}^2\left[ A_m^{*k}B_n^{k}{\rm e}^{-i\Omega_k t}
-B_m^{k}A_n^{*k}{\rm e}^{i\Omega_k t} \right]\,.
\end{eqnarray}
The matrix elements $\alpha_{mn}(t)$, $\beta_{mn}(t)$ obey the relation
\begin{equation}
\label{eq8}
\sum_{n=1}^2\left[ |\alpha_{mn}(t)|^2
-|\beta_{mn}(t)|^2 \right] = 1 \,.
\end{equation}

In the most general case it is natural to expect that at the initial stage of the
time evolution the fields can be found in the superposition of coherent and
chaotic states.
For example, the Bose condensate of ``cold" atoms can be described by
a coherent state. However, there is always a nonzero temperature which spoils
the plain coherence and brings some chaoticity (the decoherence effects).
Similar phenomenon occurs at the interaction of the laser beam with
vibrational modes in media, which are characterized by some
thermal distribution. Therefore, we assume that the initial density
matrix can be presented in the following factorized form
\beq
\hat{\varrho }(0)=
\prod_{j=1,2}\frac{\langle n_j\rangle^{\hat{b}_j^\dag\hat{b}_j}}
{(1+\langle n_j\rangle )^{\hat{b}_j^\dag\hat{b}_j+1}}.
\label{mat}
\eeq
In Eq.(\ref{mat}) we introduced new operators
$\hat{b}_{1,2}=\hat{c}_{1,2}-\alpha_{1,2}$,
where $\alpha_{1,2}$ are initial coherent amplitudes of
the fields $\hat{c}_{1,2}$ 
and the averages $\langle n_j\rangle$ are associated
with mean numbers of the bosons in the corresponding chaotic states.

To trace the time evolution of the quantum state of
the fields $\hat{c}_1$ and $\hat{c}_1$ governed by the Hamiltonian (\ref{eq1}),
it is convenient to
introduce the Wigner function, instead of the density matrix.
It can be done by dint of the symmetric characteristic
function
\begin{equation}
\label{eq9}
\chi(\mu_1,\mu_2;t)={\rm Tr}\left\{ \hat{\varrho }(0)
\exp \left[\sum_{m=1}^2\left(\mu_m\hat{c}_m^\dag(t)-\mu_m^*\hat{c}_m(t)\right)\right]\right\}.
\end{equation}
It is straightforward to show with the aid of Eq.~(\ref{eq6}) that
the Wigner function  obeys the relation
\begin{eqnarray}
\label{eq10}
&&W(\nu_1,\nu_2;t)=\frac{1}{\pi^4}\int d^2\mu_1d^2\mu_2
\chi(\mu_1,\mu_2;t)\exp{\cal D}
 \nonumber\\[1mm]
&&=\frac{1}{\pi^4}\int d^2\mu_1(t)d^2\mu_2(t)
\chi(\mu_1(t),\mu_2(t);0)
\exp{\cal D}_1\nonumber\\[1mm]
&&=W(\nu_1(t),\nu_2(t);0)\,.
\end{eqnarray}
Here ${\cal D}=\sum_{m=1}^2(\nu_m\mu_m^* -\nu_m^*\mu_m)$,
${\cal D}_1=\sum_{m=1}^2[\mu_m^*(t)\nu_m(t)-\mu_m(t)\nu_m^*(t)]$,
and $\xi_m(t)=\sum_{n=1}^2\left[ \alpha_{nm}^*(t)\xi_n
- \beta_{nm}(t)\xi_n^* \right]$ (where $\xi\!\!\equiv\!\!\mu$ or $\nu$).
It means that for initial Gaussian states the Wigner function
retains its Gaussian distribution in the process
of evolution being specified by the Hamiltonian (\ref{eq1}).

Thus, if the fields $\hat{c}_1$ and $\hat{c}_1$ are initially in the states
being superpositions of coherent and chaotic states, i.e., their Wigner
functions have the form
\beq
\label{eq11}
W_i(c_i;0)=\frac{1}{\pi (\langle n_i\rangle +1/2)}
\exp \left[ -\frac{|c_i-\alpha_i|^2}{\langle n_i\rangle +1/2} \right]
,\quad i=1,2
\eeq
the system Wigner function at time $t$ can be
described by the following expression
\beq
\label{eq12}
W(c_1,c_2;t)=\exp\left\{-\left[{\cal F}_1+{\cal F}_2\right]\right\}/\pi^2
{\cal N}
\eeq
with
\begin{eqnarray}
\label{eq13}
{\cal N}& =  & {\cal N}_1\times{\cal N}_2\equiv
(\langle n_1\rangle +1/2)\times(\langle n_2\rangle +1/2)\,,\nonumber\\[1mm]
{\cal F}_1& = &A_1(t)|c_1|^2+A_2(t)|c_2|^2-(B_1^*(t)c_1^2\nonumber\\[1mm]
&+&B_2^*(t)c_2^2+B_3^*(t)c_1c_2
-B_4^*(t)c_1c_2^*+{\rm c.c.})\,,\\
{\cal F}_2&= &\sum_{i=1}^2(|\alpha_i|^2/{\cal N}_i
+[D_i^*(t)c_i+{\rm c.c.}])\,.
\nonumber\\[1mm]
\end{eqnarray}
Here, the coefficients $A,B,D$ are defined as
\begin{eqnarray}
A_1(t)&=&\frac{|\alpha_{11}(t)|^2+|\beta_{11}(t)|^2}{{\cal N}_1}
+\frac{|\alpha_{12}(t)|^2+|\beta_{12}(t)|^2}{{\cal N}_2}\,,
\nonumber\\[1mm]
A_2(t)&=&\frac{|\alpha_{21}(t)|^2+|\beta_{21}(t)|^2}{{\cal N}_1}
+\frac{|\alpha_{22}(t)|^2+|\beta_{22}(t)|^2}{{\cal N}_2}\,,
\nonumber\\[1mm]
B_1(t)&=&\frac{\alpha_{11}(t)\beta_{11}(t)}{{\cal N}_1}
+\frac{\alpha_{12}(t)\beta_{12}(t)}{{\cal N}_2}\,,
\nonumber\\[1mm]
B_2(t)&=&\frac{\alpha_{21}(t)\beta_{21}(t)}{{\cal N}_1}
+\frac{\alpha_{22}(t)\beta_{22}(t)}{{\cal N}_2}\,,
\nonumber\\[1mm]
B_3(t)&=&\frac{\alpha_{11}(t)\beta_{21}(t)+\alpha_{21}(t)\beta_{11}(t)}
{{\cal N}_1}\\[1mm]
&+&\frac{\alpha_{12}(t)\beta_{22}(t)+\alpha_{22}(t)\beta_{12}(t)}
{{\cal N}_2}\,,
\nonumber\\[1mm]
B_4(t)&=&\frac{\alpha_{11}(t)\alpha_{21}^*(t)+\beta_{11}(t)\beta_{21}^*(t)}
{{\cal N}_1}\nonumber\\[1mm]
&+&\frac{\alpha_{12}(t)\alpha_{22}^*(t)+\beta_{12}(t)\beta_{22}^*(t)}
{{\cal N}_2}\,,
\nonumber\\[1mm]
D_1(t)&=&\frac{\alpha_{11}(t)\alpha_1-\beta_{11}(t)\alpha_1^*}
{{\cal N}_1}
+\frac{\alpha_{21}(t)\alpha_2-\beta_{21}(t)\alpha_2^*}
{{\cal N}_2}\,,
\nonumber\\[1mm]
D_2(t)&=&\frac{\alpha_{12}(t)\alpha_1-\beta_{12}(t)\alpha_1^*}
{{\cal N}_1}
+\frac{\alpha_{22}(t)\alpha_2-\beta_{22}(t)\alpha_2^*}
{{\cal N}_2}\nonumber\,.
\end{eqnarray}

The Gaussian character of the system state (\ref{eq12}) is well
suited to introduce the entanglement measure within the approach
proposed in Ref.\onlinecite{VV02}.
The measure of entanglement between the fields
can be calculated analytically in the form of the logarithmic negativity
via the symplectic spectrum of the partial transpose of the covariance
matrix. To proceed along this line we present the bilinear
in fields part of the Wigner function (\ref{eq12})
by dint of the relation
${\cal F}_1=\frac{1}{2}{\bf \zeta}^T{\bf V}^{-1}{\bf \zeta}$.
Here, the inverse variance matrix of the form
\begin{equation}
{\bf V}^{-1}=\left(
\begin{array}{cc}
{\bf X}_1   & {\bf Y} \\
{\bf Y}^T & {\bf X}_2
\end{array}
\right)\
\end{equation}
is determined by the matrices
\begin{eqnarray}
{\bf X}_i& = &\left(
\begin{array}{cc}
A_i-2{\rm Re}B_i  & -2{\rm Im}B_i\\
-2{\rm Im}B_i      & A_i+2{\rm Re}B_i
\end{array}
\right),\,\, i=1,\,2\,,\\
{\bf Y}& = &\left(
\begin{array}{cc}
{\rm Re}(B_4-B_3) & -{\rm Im}(B_4+B_3)\\
-{\rm Im}(B_3-B_4) & {\rm Re}(B_4+B_3)
\end{array}
\right).
\end{eqnarray}
And ${\bf \zeta}^T=(q_1,p_1,q_2,p_2)$ is a transposed four-vector
whose elements are the quadrature-component variables defined as
$c_i=(q_i+ip_i)/{\sqrt{2}}$.

\begin{figure}[ht]
\includegraphics[height=0.16\textheight,width=0.28\textwidth,clip]{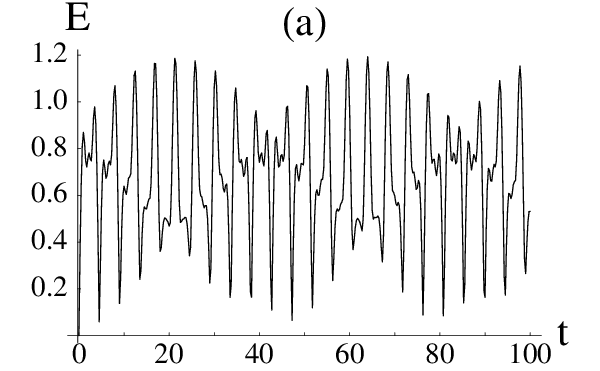}\\[0.3cm]
\includegraphics[height=0.16\textheight,width=0.28\textwidth,clip]{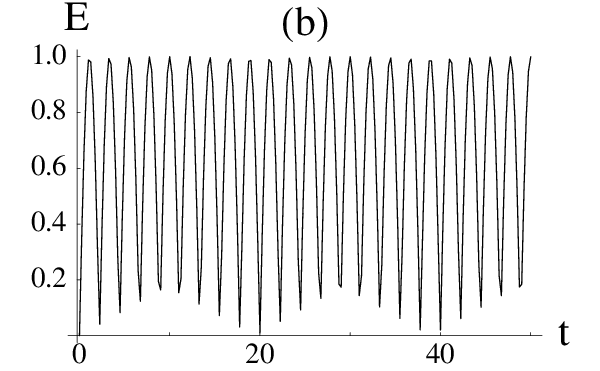}
\caption{\label{fig1}
Entanglement
Eq.~(\protect\ref{mes}) between the modes being initially in coherent states.
Model parameters are
$\omega_1=1$, $\omega_2=2$ and (a) $g_1=g_2=0.5$, (b) $g_1=0$, $g_2=0.5$.
}
\end{figure}

As a result, the logarithmic negativity
\begin{equation}
\label{mes}
E=-\frac{1}{2}\log_2(4{\cal A})\,,
\end{equation}
where
\begin{eqnarray}
{\cal A}& = &{\cal B}-\sqrt{{\cal B}^2-\det {\bf V}}\,,\\
{\cal B}& = &\frac{\det {\bf {\tilde  X}}_1+\det {\bf {\tilde  X}_2}}{2}-\det
{\bf {\tilde Y}}\,,
\end{eqnarray}
determines the strength of the entanglement for $E>0$. 
Here, ${\bf {\tilde X}_i}$ and ${\bf {\tilde Y}}$ are diagonal and nondiagonal
block matrices of the variance matrix ${\bf V}$, respectively.
For $E\leq0$
the composite state is  separable according to the
Peres-Horodecki criterion \cite{per} (see also Refs.\cite{sim}).
This measure enables us
to trace the evolution of non-classical correlations in the system and
the conditions for their amplification.

The considered model
provides various possibilities
to study cumulative effects produced by
a different strength of coupling between the fields
prepared in the initial states of different degree of chaoticity (different
temperatures).
In the present paper we discuss a few interesting cases
of the dynamical control of the entanglement by dint of different
choices of the initial states and coupling constants. For illustration
we consider the case $\omega_1=1$,
$\omega_2=2$ (in relative units). Note, that at a given energy $\hbar\omega_i$
of the $\hat c_i$-field initial state, the temperature $T_i$
fixes the mean number of bosons $\langle n_i \rangle$, i.e.,
$\langle n_i \rangle =\left\{\exp\left[\hbar\omega_i /(k_{\rm B}T_i)\right]-
1\right\}^{-1}$. All frequencies and coupling constants are determined in 
units of the frequency $\omega_1$ and the time is scaled by 
inversed $\omega_1$.

At zero temperature, when both the fields are in coherent (vacuum) initial
states, the interaction with equal coupling constants ($g_1=g_2$)
produces the entangled system state (see Fig.~1a).
However, the degree of entanglement is not monotonic with a time.
It displays an oscillatory character with the distinctive alteration
of the entanglement maxima and minima that can be considered to be revivals and
collapses. In this regime  the fields $\hat c_1$ and $\hat c_2$ were shown to
be evolved into squeezed states \cite{CN90} that is a consequence of such
quantum correlations. In case the system is associated with the interaction of
a light with ion oscillations in a crystal (optical phonons),
a suppression of zero-point optical vibrations (squeezing) may
affect the electron-pnonon interaction in nanostructures.
We speculate that
the nonmonotonic behavior of the entanglement and the squeezing
results in the interaction strength which would fluctuate in time.
Note that this interaction being essential to the understanding of
electron-spin decoherence effects in quantum dots  is
assumed to be independent on time (cf \cite{sem}).

\begin{figure}[ht]
\includegraphics[height=0.16\textheight,width=0.28\textwidth,clip]{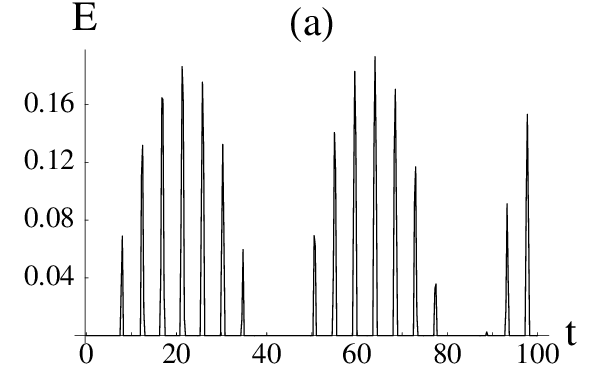}\\[0.3cm]
\includegraphics[height=0.16\textheight,width=0.28\textwidth,clip]{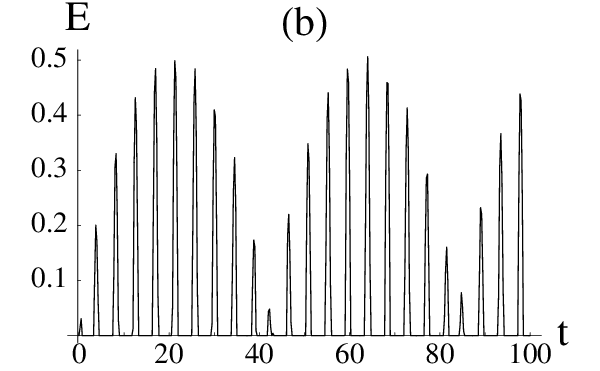}
\caption{\label{fig2}
The same as in Fig.~\ref{fig1} for (a) chaotic states with
$\langle n_1\rangle =\langle n_2\rangle =0.5$ and
(b) coherent and chaotic states when $\langle n_2\rangle =5$.
}
\end{figure}

The entanglement also occurs
for the interaction regime $g_1=0$ and
$g_2\neq 0$ ($k_{\rm B}T=0$). In this case the interaction produces
the picket-fence effect for the entanglement
of equal maxima without revivals (see Fig.~1b)
but with the periodical alteration of its minima.
However, in this regime the fields $\hat c_1$ and $\hat c_2$
remain in coherent states, i.e., the squeezing does
not take place \cite{CN90}. It appears that
the interaction that creates (annihilates) the boson pair
(associated with the strength $g_2$) is
enough to produce the entanglement. On the other hand,
both interactions (associated with the strengths $g_1$ and $g_2$)
are indispensable in the Hamiltonian (\ref{eq1}) in order to
produce the squeezing of the coherent initial states.
The interaction of the $g_1\neq 0$ and $g_2=0$ type
does not yield an entangled state for the system at all,
as well as the single-mode squeezing for the coherent states.

Thermal fluctuations are expected to attenuate the entanglement.
In particular, when both the fields are initially in chaotic states,
the degree of their entanglement in the regime of equal coupling constants
becomes very small (see Fig.~2a).
Moreover, there is a critical temperature
above which the entanglement disappears
and the system becomes separable
(for equal coupling constants in our choice of the
initial energies $\omega_{1,2}$ it corresponds to
$\langle n_{1,2}\rangle \approx 0.63$).
However, if one of the initial states is
found in a coherent state while the other is in a chaotic one,
the entanglement evolves at regular intervals with
a periodical alteration of maxima (see Fig.~2b).
With a proper combination of the initial energies of the system
and a choice of number of bosons in the chaotic (initial) state,
one may maintain the entanglement at a noticeable level,
even at high temperatures.

In conclusion, the considered two-mode boson
model demonstrates that the entanglement can be
controlled in time by appropriate choice of the interaction at different temperature
regimes. The obtained results show that this plain system exhibits
a rich dynamics from the view point of the quantum information
theory and its possible physical applications.

\section*{Acknowledgements}
We  thank Montserrat Casas and Antoni Borras
for usefull discussions.
This work was partly supported by
Grant No. FIS2005-02796 (MEC, Spain) and
RFBR Grant No. 08-02-00118 (Russia).

\end{document}